\documentclass[a4paper,fleqn]{cas-dc}
\usepackage[authoryear]{natbib}
\setcitestyle{aysep={}} 
\usepackage{siunitx}
\usepackage{bm}

\usepackage{bigints}
\usepackage{ulem}
\usepackage{hyperref, bookmark, tocloft}

\usepackage{indentfirst}
\usepackage{caption}
\usepackage{float}
\usepackage{color}
\usepackage{threeparttable}
\usepackage{multicol}
\usepackage{multirow}
\usepackage{tabularx}
\usepackage{longtable}
\usepackage{geometry}
\usepackage{booktabs}
\usepackage[noend]{algpseudocode}
\usepackage{amsthm}
\usepackage{algorithmicx,algorithm}

\theoremstyle{plain}

\newtheorem{theorem}{Theorem}
\newtheorem{corollary}{Corollary}
\newtheorem{example}{Example}

\def\E{\operatorname*{E}}
\def\Var{\operatorname*{Var}}

\def \w{\operatorname*{\mathbf{w}}}

\def \u{\operatorname*{\mathbf{u}}}

\def \bsf  {\boldsymbol}
\def \mbf {\mathbf}

\usepackage{fancyhdr}
\makeatletter
\newcommand{\thickhline}{%
    \noalign {\ifnum 0=`}\fi \hrule height 1pt
    \futurelet \reserved@a \@xhline
}

\newcolumntype{"}{@{\hskip\tabcolsep\vrule width 1pt\hskip\tabcolsep}}
\makeatother

\makeatletter
\newcommand\myemail[3]{
\edef\@tempa{mailto:#1 }%
\edef\@tempb{\expandafter\html@spaces\@tempa\@empty}%
\href{\@tempb}{#3}}

 \catcode\%=11
 \def\html@spaces#1 #2{#1\ifx#2\@empty\else\expandafter\html@spaces\fi#2}
 \catcode\%=14
\makeatother

\begin{document}
\let\WriteBookmarks\relax
\def\floatpagepagefraction{1}
\def\textpagefraction{.001}
 \title[mode = title]{Estimating Conditional Average Treatment Effects with Heteroscedasticity by Model Averaging and Matching}{}
\shorttitle{\normalfont{}}
\author[1]{Pengfei Shi}[style=chinese]
\author[2]{Xinyu Zhang}[style=chinese]
\cormark[1]
\author[3]{Wei Zhong}[style=chinese]

\address[1]{Paula and Gregory Chow Institute for Studies in Economics, Xiamen University, Fujian, China}

\address[2]{Academy of Mathematics and Systems Science, Chinese Academy of Sciences, Beijing, China}

\address[3]{MOE Key Lab of Econometrics, WISE and Department of Statistics and Data Science in SOE, Xiamen University, Fujian, China}

\shortauthors{P.Shi, W.Zhong, X.Zhang}  

\begin{abstract}
	We propose a model averaging approach, combined with a partition and matching method to estimate the conditional average treatment effects under heteroskedastic error settings. The proposed approach has asymptotic optimality and consistency of weights and estimator. Numerical studies show that our method has good finite-sample performances.
\end{abstract}
\begin{keywords}
	Asymptotic optimality; Cross validation; Heteroscedasticity; Model averaging; Treatment effects
\end{keywords}

\maketitle
 
\section{Introduction}
Causal inference is essential to empirical investigations, in which the estimation of treatment effects on the outcome is a key point. \cite{rolling2014model} proposed a treatment effect cross validation (TECV) method to select the model within a candidate model set that is globally the most accurate for estimating the treatment effects. Instead of selecting one specific model, we consider a model averaging approach, which incorporates model uncertainty into the estimation process. There are lots of literature to select the weights towards some form of optimality. \cite{hansen2007least} and \cite{wan2010least} proposed least squares model averaging by Mallow's criterion. \cite{hansen2012jackknife} and \cite{zhang2013model} developed jackknife model averaging. \cite{lu2015jackknife} utilized the idea of jackknife model averaging in quantile regressions. \cite{xie2015prediction} proposed prediction model averaging (PMA) estimator. \cite{liu2018averaging} proposed averaging estimators for kernel regressions. \cite{feng2020sparsity} showed that the weights vector obtained by Mallow's model averaging method has a sparisty property. \cite{kitagawa2016model} minimised the approximated mean squared error of a semiparametric estimator in terms of treatment effects based on a model averaging method. \cite{antonelli2020averaging} proposed averaging causal estimators in high dimensions which merges multiple models in different frameworks without theoretical guarantees. \cite{rolling2019combining} developed treatment effect estimation by mixing (TEEM), which has a risk bound for the estimator without asymptotic optimality. \cite{zhao2023model} proposed a model averaging approach to estimate CATE, which has asymptotic optimality under homoscedasticity.  

In this paper, we consider the problem of estimating treatment effects on the response variable under heteroskedastic error settings. In our weights choice criterion, we firstly use the partition and matching method introduced by \cite{rolling2014model} to approximate the true treatment effects, then minimize the mean squared error of leave-one-out cross validation, which is also known as the jackknife method. Based on the theory developed in \cite{li1987asymptotic} and \cite{wan2010least}, we derived the asymptotic optimality under heteroskedastic error settings. Furthermore, we also derived the consistency of weights and estimator for CATE, which has not been studied in the aforementioned articles on estimating treatment effects using model averaging. Numerical studies show that our method has good finite-sample performance.

The remainder of this paper is organized as follows. Section 2 introduces the model framework, the proposed model averaging estimator for CATE and its asymptotic properties. Numerical studies are presented in Section 3. The proofs of the theorem and some results of numerical studies are presented in the supplementary.

\section{Methodology}\label{section2}
\subsection{Model Framework}
We consider a general regression framework in which the response $Y$ is dependent on a binary treatment variable $T\in \{t,c\}$ and some baseline covariates $\mbf{u}\in \mathbb{R}^p$. The data generating process can be written as follows:
\begin{equation}
	\label{DGP}
	Y_i=[f_t(\mbf{u}_i)+\zeta_i]I(T_i=t)+[f_c(\mbf{u}_i)+\nu_i]I(T_i=c),
\end{equation}
where $\mbf{u}_i$ are independent and identically distributed (IID) from some unknown distribution $P_{\mbf{U}}$ with support $\mathcal{U}\subset \mathbb{R}^p$, and $f_t(\cdot)$ and $f_c(\cdot)$ are the expectation of $Y_{i,t}$ and $Y_{i,c}$ given the covariates respectively, where $Y_{i,t}$ and $Y_{i,c}$ are the potential outcomes under treatment and control, respectively. The random errors are $\zeta_i$ and $\nu_i$, where we assume a heteroskedastic error setting, that is, $\E(\zeta_i| \mbf{u}_i)=\E(\nu_i| \mbf{u}_i)=0$, $\E(\zeta_i^2| \mbf{u}_i)=\sigma_{i,t}^2$ and $\E(\nu_i^2| \mbf{u}_i)=\sigma_{i,c}^2$.
Our goal is to estimate $\text{CATE}(\mbf{u}):=\E[(Y_{i,t}-Y_{i,c})| \mbf{u}_i=\mbf{u}]$. Since we can not obesrve both $Y_{i,t}$ and $Y_{i,c}$ simultaneously, to make this problem identifiable, we need the unconfounded assignment, which is that $\{Y_{i,t},Y_{i,c}\}\perp T_i| \mbf{u}_i$. This is the basic assumption needed in lots of literature on causal inference based on observational study and as we collect more and more features, this assumption becomes easier to satisfy (\cite{rolling2014model,antonelli2020averaging,fan2022estimation}). Under the unconfounded assignment, we have that 
\begin{equation}
	\text{CATE}(\mathbf{u})=f_t(\u)-f_c(\u):=\Delta(\u).
\end{equation} 
In this work, we focus on estimating $\Delta(\mbf{u})$ because it is identifiable in most experiments and observational studies and we will refer to $\Delta(\mbf{u})$ as the treatment effect during the remainder of this paper.

We consider $K$ candidate models, which utilize different $K$ subsets of $\mbf{u}_i$. Denote the $i$-th vector of the covariates matrix in the $k$-th candidate model as $\mbf{u}_i^k=(u_{1i}^k,...,u_{p_ki}^k)^T$, where $p_k$ is the number of covariates used in the $k$-th candidate model. For simplicity, we consider a linear model, then the $k$-th candidate model can be written by
\begin{equation}
	Y_i=[\mathbf{u}_i^{k^T}\boldsymbol{\beta}^k+\zeta_i]I(T_i=t)+[\mathbf{u}_i^{k^T}\boldsymbol{\gamma}^k+\nu_i]I(T_i=c),
\end{equation}
for $i=1,...,n$, where $\bsf{\beta}^k$ and $\bsf{\gamma}^k$ are the parameter in the $k$-th model to be estimated. Let $\mbf{Y}_{ta}$ be the response vector in the treatment group with length $n_t$ and $\mbf{Y}_{ca}$ be the response vector in the control group with length $n_c$. Denote the covariates matrix corresponding to treatment and control groups as $\mbf{U}_{ta}^k$ and $\mbf{U}_{ca}^k$ respectively. We assume that $\mbf{U}_{ta}^k$ and $\mbf{U}_{ca}^k$ are of full column rank. Our primary focus will be on least-square estimators, in which case,
$
	\hat{\bsf{\beta}}^k=(\mbf{U}_{ta}^{k^T}\mbf{U}_{ta}^k)^{-1}\mbf{U}_{ta}^{k^T}\mbf{Y}_{ta},~\hat{\bsf{\gamma}}^k=(\mbf{U}_{ca}^{k^T}\mbf{U}_{ca}^k)^{-1}\mbf{U}_{ca}^{k^T}\mbf{Y}_{ca}.
$
Then the estimator of the treatment effect $\Delta(\mbf{u}_i)$ under the $k$-th candidate model is the difference in the regressed functions between treatment and control groups,
$
	\hat{\Delta}^k(\mbf{u}_i)=\mbf{u}_i^{k^T}\hat{\bsf{\beta}}^k-\mbf{u}_i^{k^T}\hat{\bsf{\gamma}}^k.
$
Thus, we assign the $k$-th candidate model with weight $w_k$ and the model averaging estimator can be written as
\begin{equation}
	\label{mae}
	\hat{\Delta}_i(\w)=\sum\nolimits_{k=1}^K w_k\hat{\Delta}^k(\mbf{u}_i),
\end{equation}
where $\w=(w_1,...,w_K)^\prime$ is a weights vector belonging to the set
$
	Q_n=\left\{\w\in[0,1]^K:\sum_{k=1}^Kw_k=1\right\}.
$
We will discuss that how to choose the weights vector $\w$ in the next subsection. In this work, we need that the candidate models are all linear models, which are estimated by least square method to demonstrate asymptotic properties. In this regard, our approach is more restrictive than the method of \cite{rolling2019combining}.

\subsection{Weights Choice Criterion}
We utilize the “partition and match” idea in \cite{rolling2014model}. In specific, we partition the feature space into lots of cells. Without loss of generality, we let the support of the probability density of $\mbf{u}$ be $[0,1]^p$. Then we partition the feature space into $M$ cells with side length $h$. Next, in the $m$-th cell $(m=1,...,M)$, we can randomly choose a pair of observations $(m,m^*)$ such that $T_m=t$ and $T_{m^*}=c$. Denote $\mbf{u}_m^t$ and $\mbf{u}_m^c$ as the covariate vectors corresponding to the treatment observation and control observation in the $m$-th pair, respectively. Define $b_m=f_c(\mbf{u}_m^t)-f_c(\mbf{u}_m^c)$. Let $Y_m^t,Y_m^c,\zeta_m^t,\nu_m^c$ be the corresponding responses and errors respectively, and $\eta_m=\zeta_m^t-\nu_m^c$. It is reasonable to approximate $\Delta({\mbf{u}_m^t})$ by $Y_m^t-Y_m^c$, which has an approximation bias term $b_m$ and an error term $\eta_m$ additionally. Furthermore, if we use the side length $h$ suggested in \cite{rolling2014model}, the approximation bias $b_m,m=1,...,M$ have a uniform bound, which provides insurance for the accuracy of our estimator.

In this paper, we propose a jackknife selection of $\mbf{w}$, which is also known as leave-one-out cross validation. After choosing $M$ pairs of observations, we define that $\bsf{\Delta}=(\Delta(\mbf{u}_1^t),...,\Delta(\mbf{u}_M^t))^\prime$ and the $k$-th jackknife estimator for $\bsf{\Delta}$ is $\tilde{\bsf{\Delta}}^k=(\tilde{\Delta}^k(\mbf{u}_1^t),...,\tilde{\Delta}^k(\mbf{u}_M^t))^\prime$, where $\tilde{\Delta}^k(\mbf{u}_i^t)$ is the estimator $\hat{\Delta}^k(\mbf{u}_i^t)$ computed with the $i$-th observation deleted. We take average on the $K$ jackknife estimators and obtain that $\tilde{\bsf{\Delta}}(\mbf{w})=\sum_{k=1}^K w_k \tilde{\bsf{\Delta}}^k$. Let $\tilde{Y}_m=Y_m^t-Y_m^c$ and $\tilde{\mbf{Y}}=(\tilde{Y}_1,...,\tilde{Y}_M)^\prime$. Then we propose the weights choice criterion based on the residuals of the jackknife averaging estimator:
\begin{equation}
	\label{criterion}
	\hat{\mathbf{w}} = \mathop{\arg\min}_{\w\in Q_n}~\text{CV}(\w) =\mathop{\arg\min}_{\w\in Q_n}~ ||\tilde{\bsf{\Delta}}(\w)-\tilde{\mbf{Y}}||^2.
\end{equation}
We take this chosen weights to \eqref{mae} and obtain the jackknife model averaging estimator of the treatment effects $\hat{\Delta}_i(\w)=\sum_{k=1}^K \hat{w}_k\hat{\Delta}^k(\mbf{u}_i)$ and we label our method as JMA.

\subsection{Asymptotic Properties}
In this subsection, we study the theoretical properties of the proposed estimator.
Let $\hat{\bsf{\Delta}}^k=(\hat{\Delta}^k(\mbf{u}_1^t),...,\hat{\Delta}^k(\mbf{u}_M^t))^\prime$, $\hat{\bsf{\Delta}}(\mbf{w})=\sum_{k=1}^Kw_k \hat{\bsf{\Delta}}^k$, and $\bsf{\Delta}=(\Delta(\mbf{u}_1^t),...,\Delta(\mbf{u}_M^t))^\prime$. Define the squared error of $\hat{\bsf{\Delta}}(\w)$ as $L_n(\mbf{w})=||\hat{\bsf{\Delta}}(\mbf{w})-\bsf{\Delta}||^2$ and the corresponding conditional risk as $R_n(\mbf{w})=\E[L_n(\mbf{w})\mid \mbf{u}]$. We then give the asymptotic optimality. 
\begin{theorem}[Asymptotic Optimality]
	Suppose that conditions (3.1)-(3.4) in the supplementary hold. Then our averaging estimator has the asymptotic optimality in the sense of mean squared error as $n\rightarrow \infty$,
	\begin{equation}
		\centering
	\label{OPT}
	\frac{L(\hat{\mathbf{w}})}{\mathop{\inf}_{\mathbf{w}\in Q_n}L(\mathbf{w})}\rightarrow 1.
	\end{equation}
\end{theorem}

\noindent{The conditions and the corresponding remarks and proofs are presented in the Section 3 in the supplementary. This property guarantees that when all the candidate models are misspecified, the proposed estimator can achieve the possible squared error asymptotically. For the number of candidate models $K$, we need that $K \cdot p^{1 / 2} \cdot n^{1 / 2}/\tilde{\xi}_n \rightarrow 0$. Assume that the order of $\tilde{\xi}_n$ is $n^{1-\delta}$ and $\delta<1/2$, then $K$ and $p$ are allowed to grow to infinity with order $K\cdot p^{1/2}=o_p(n^{1/2-\delta})$.}

Next, we consider the case, where the candidate model set contains the true model. Let the subset $S\subseteq \{1,...,K\}$ includes the indices of the correct models, and $|S|=k_0\le K$. Without loss of generality, we can assume that the first $k_0$ models are correct. Denote $\hat{w}_\Delta=\sum_{k=1}^{k_0} \hat{w}_k$ as the sum of the weights of all the correct candidate models. 

\begin{theorem}[Consistency of weights]
	Suppose Conditions (3.5)-(3.8) in the supplementary hold. Then, as $n\rightarrow \infty$
	\begin{equation}
  \vspace{-0.2cm}
	\label{consistent}
	\hat{w}_\Delta \overset{p}{\longrightarrow} 1.
  \vspace{-0.5cm}
	\end{equation}
\end{theorem}

\noindent{The conditions and the corresponding remarks and proofs are presented in the Section 3 in the supplementary. Theorem 2 guarantees that when there exist correct models in the candidate model set, our method will assign all the weights to the correct models asymptotically. Therefore, these two properties comprehensively consider all the situations for the candidate model set and provide corresponding theoretical guarantees. Furthermore, we also derive the consistency of estimator for CATE.}

\begin{corollary}[Consistency of estimator for CATE]
    Under Conditions (3.5)-(3.8) in the supplementary, we have that
    \begin{equation}
        |\hat{\Delta}_i(\hat{\mathbf{w}})-\Delta_i|^2=o_p(1).
    \end{equation}
\end{corollary}

\noindent{This corollary shows that our proposed model averaging estimator for CATE converges to the true CATE if there exists at least one correct specified candidate model, and the proofs are in the supplementary.}

\section{Numerical Studies}\label{section3}
We compare the performances of these approaches in finite sample case including our proposed jackknife model averaging estimator, which is labeled as JMA, the AIC and BIC model selection estimators, which select the model with the smallest AIC and BIC scores respectively, smoothed-AIC (SAIC), smoothed-BIC (SBIC) estimators proposed by \cite{buckland1997model}, which assign the weights based on the AIC and BIC scores respectively, TECV model selection method, which is proposed by \cite{rolling2014model}, TEEM proposed by \cite{rolling2019combining}. In addition, we also consider the single model containing all covariates, estimated by penalized least square (PLS) with $L_1$ penalty.

\subsection{Monte Carlo Simulation}

\begin{example}
Consider the model framework in \eqref{DGP}, where
\begin{align}
	&f_t(\mathbf{u}_i)=c\cdot(0.5u_{i1}^2+0.5u_{i2}+0.5u_{i1}+0.5u_{i2}^2)\nonumber\\
&	f_c(\mathbf{u}_i)=c\cdot(0.5u_{i1}^2+0.5u_{i2})
\end{align}
where $c$ is a constant. We can vary $c$ to obtain the different values of $R^2 = [\Var(Y_i)-\Var\{T_i\cdot \zeta_i+(1-T_i)\cdot \nu_i\}]/\Var(Y_i)$, which represents the signal strength. The smaller value of $R^2$, the lower signal level. $T_i\sim \text{Bin}(1,0.5)$ is a binary treatment variable and $(u_{i1},u_{i2},u_{i3},u_{i4})^\prime \sim \mathbf{N}_{4}(\mathbf{0},\mathbf{\Sigma})$, where the $k,j$-th element of $\mathbf{\Sigma}$ is set to $\rho^{|k-j|}$, $\rho=0,0.5$. The errors are generated in the following ways:
\begin{itemize}
	\item [(1)] Design 1: $\zeta_i\sim N(0,9u_{i2}^2),~v_i\sim N(0,9u_{i2}^2)$,
	\item [(2)] Design 2: $\zeta_i\sim N(0,16u_{i2}^2),~v_i\sim N(0,4u_{i1}^2)$,
\end{itemize}
which both represent a heteroskedastic error setting. We use four candidate models listed in Table 1 in the supplement material, which are all misspecified.
\end{example}
\noindent{ We compare different methods' performances through the average squared errors }
\begin{equation*}
	\text{ASE}_j=\frac{1}{N_{new}}\sum\nolimits_{i=1}^{N_{new}}(\hat{\Delta}_n(\mbf{u}_i)-\Delta(\mbf{u}_i))^2,
\end{equation*}
where $N_{new}=10^6$ is the number of the evaluation observations. We replicate each model $J=100$ times and take the average. The simulation results are presented after normalization, that is in each replication we divided the risk by the risk of the infeasible optimal estimator.

We set the sample size $n=200,400,800$ and vary $R^2=0.3,0.4,0.5,...,0.9$. We show one result in Figure 1 and the remaining simulation results are presented in the Section 1 of the supplementary. From the results, we can see that for both design 1 and design 2, no matter $\rho$ equals to 0 or 0.5, our proposed estimator JMA performs best. Especially in the case of small samples, JMA has significant advantages over other methods. In other hand, when $R^2$ increases, the performance gap between JMA and other methods becomes large. Therefore, our proposed estimator has good performance in finite sample case in terms of average squared errors.
\vspace{-0.4cm}
\begin{figure}[ht]
\setlength{\abovecaptionskip}{-0.2cm}
	\centering
		\includegraphics[width=0.35\textwidth]{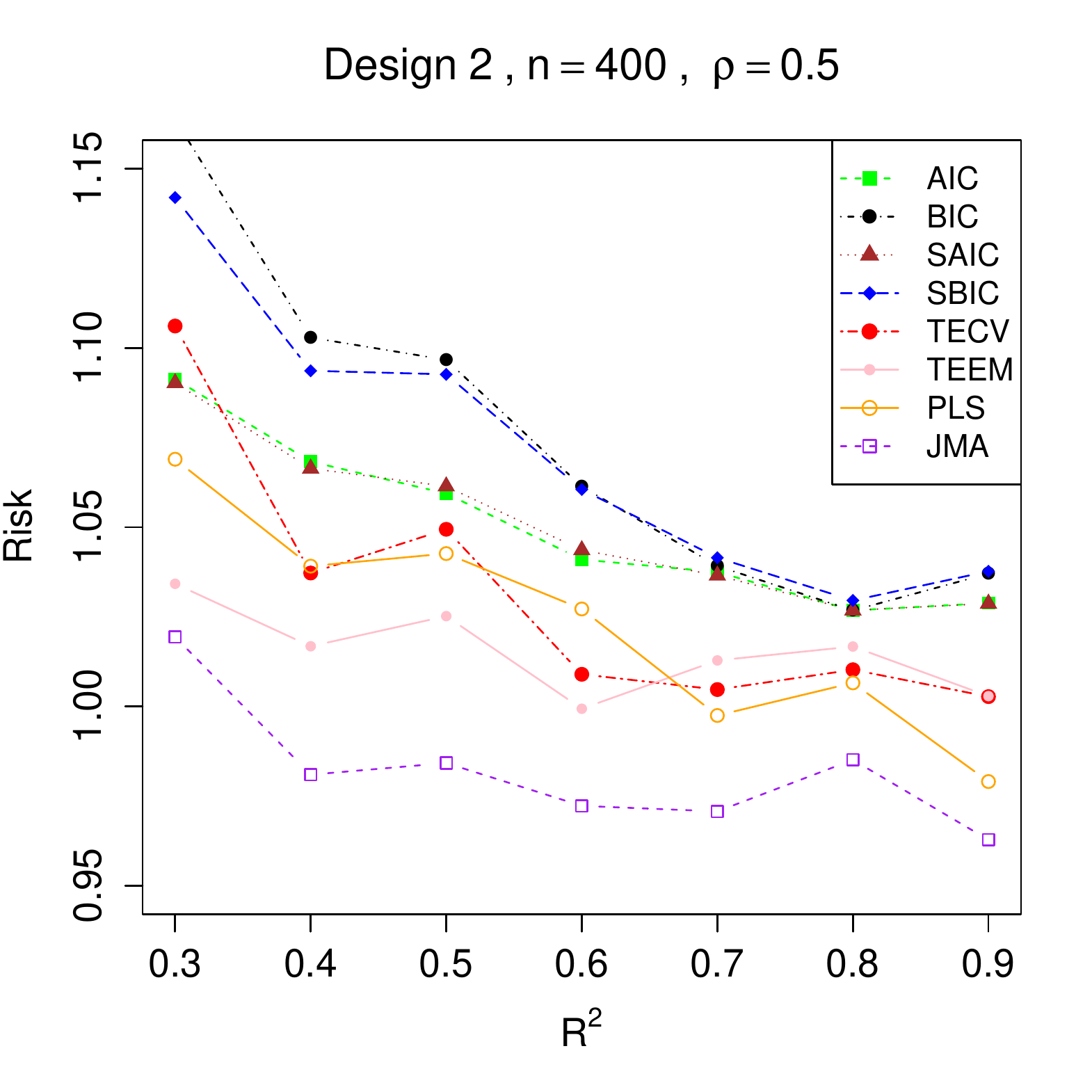}
  \vspace{-0.5cm}
		\caption{Result of simulations for Example 1}
\end{figure}
\vspace{-0.7cm}
\begin{example}
	We use the same data generating process as Example 1, and the four candidate models are listed in Table 2 in the supplement material, where the first candidate model is correct and others are misspecified.
\end{example}
\noindent{In the j-th replication, we record the weight assigned to the first model, which is the correct model, $\hat{w}_{\Delta,j}=\hat{w}_1$.}
We replicate each model $J=100$ times and take the average. We show one result in Figure 2 and the remaining results are presented in the Section 1 of the supplementary. From the results, we can see that in all the settings, as the sample size increases, the weight assigned to the correct model also increases and becomes closer to 1. This result supports our theoretical result about weights consistency.
\begin{figure}[ht]
	\centering
		\includegraphics[width=0.35\textwidth]{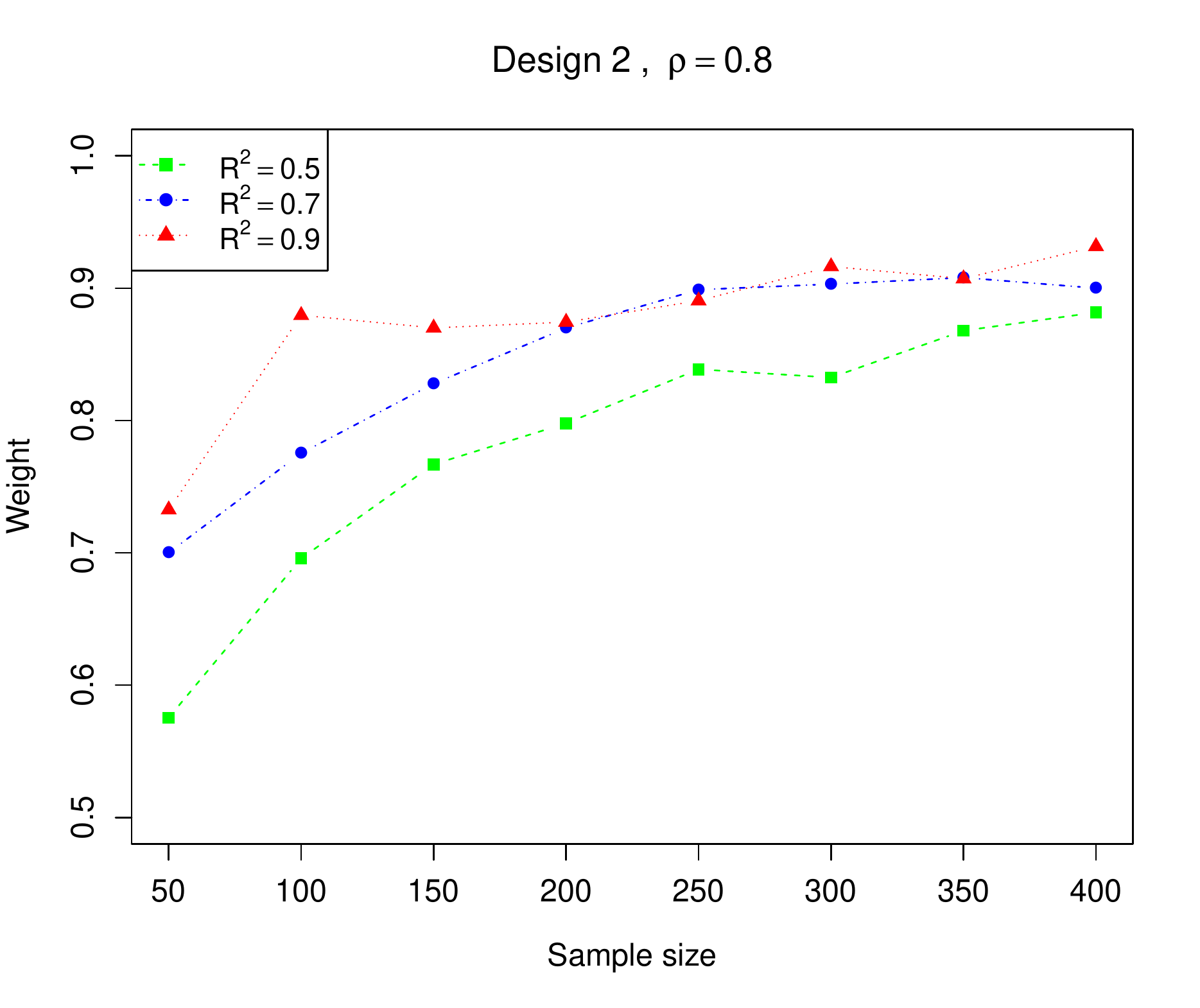}
  \vspace{-0.5cm}
		\caption{Result of simulations for Example 2}
  \vspace{-0.7cm}
\end{figure}

\subsection{Empirical study on NSWD data set}
In our empirical application, we consider the \cite{lalonde1986evaluating} National Supported Work Demonstration data set. The aim of this analysis is to estimate the treatment effects of work training on the outcome. The data set contains 722 observations, of which 297 are in the treatment group and 425 are in the control group. We consider the difference in the square root of income between 1975 (pre-treatment) and 1978 (post-treatment) as our response. The treatment variable $T$ equaling to 1 means this person accepted training in the NSW Demonstration and 0 otherwise. We use four baseline covariates (square root of 1975 income, age, years of education, marital status), which are all measured before the training to identify heterogeneous treatment effects. We take the interaction terms between covariates and quadratic terms of 3 continous variables to construct totally 25 candidate models. We list all the covariates each candidate model uses in the Table 3 in the Section 2 in the supplementary.

We evaluate the different estimators by a “guided simulation” experiment, that is, we choose the 19-th model, the 5-th and 25-th model, which are selected by three model selection methods AIC, BIC, TECV respectively as the “true” processes. For each “true” process, we use this model to estimate error variance, which is denoted as $\hat{\sigma}^2$. Then for each replication, we generate random variables $u_i$ from $\text{Unif}(0.5,1.5)$, and use $u_i^2*\hat{\sigma}^2$ as the variance to generate a noise variable from zero-mean Gaussian distribution. Then we generate the same number of observations of $Y$ by augmenting the estimated regression using this “true” model with the noise variables. For each method, we can obtain the estimated treatment effects $\hat{\Delta}(\mbf{u}_i)$ and we evaluate the performance of the method by the average of the squared errors $(\Delta(\mbf{u}_i)-\hat{\Delta}(\mbf{u}_i))^2$ across the 722 observations. In each replication, we can obtain the “true” treatment effects $\Delta(\mbf{u}_i)$ from the assumed true process. We repeat the experiment 100 times.

We show one result in Figure 3 and the remaining results are presented in the Section 2 of the supplementary. From the results we can see that when we assume the 19-th or 25-th candidate model as the “true” process, the JMA estimator has the best performance among all the methods and also results in the smallest variance of average squared errors. When we use the 5-th candidate model as the “true” process, the BIC and SBIC methods perform best since we choose the candidate model determined by BIC criterion as the “true” process. In this case, the JMA estimator still has a good performance. Therefore, our proposed method has a good performance in real data analysis in terms average squared errors and the estimator has small variance thus it is robust.
\begin{figure}[ht]
	\centering	
	\includegraphics[width=0.35\textwidth]{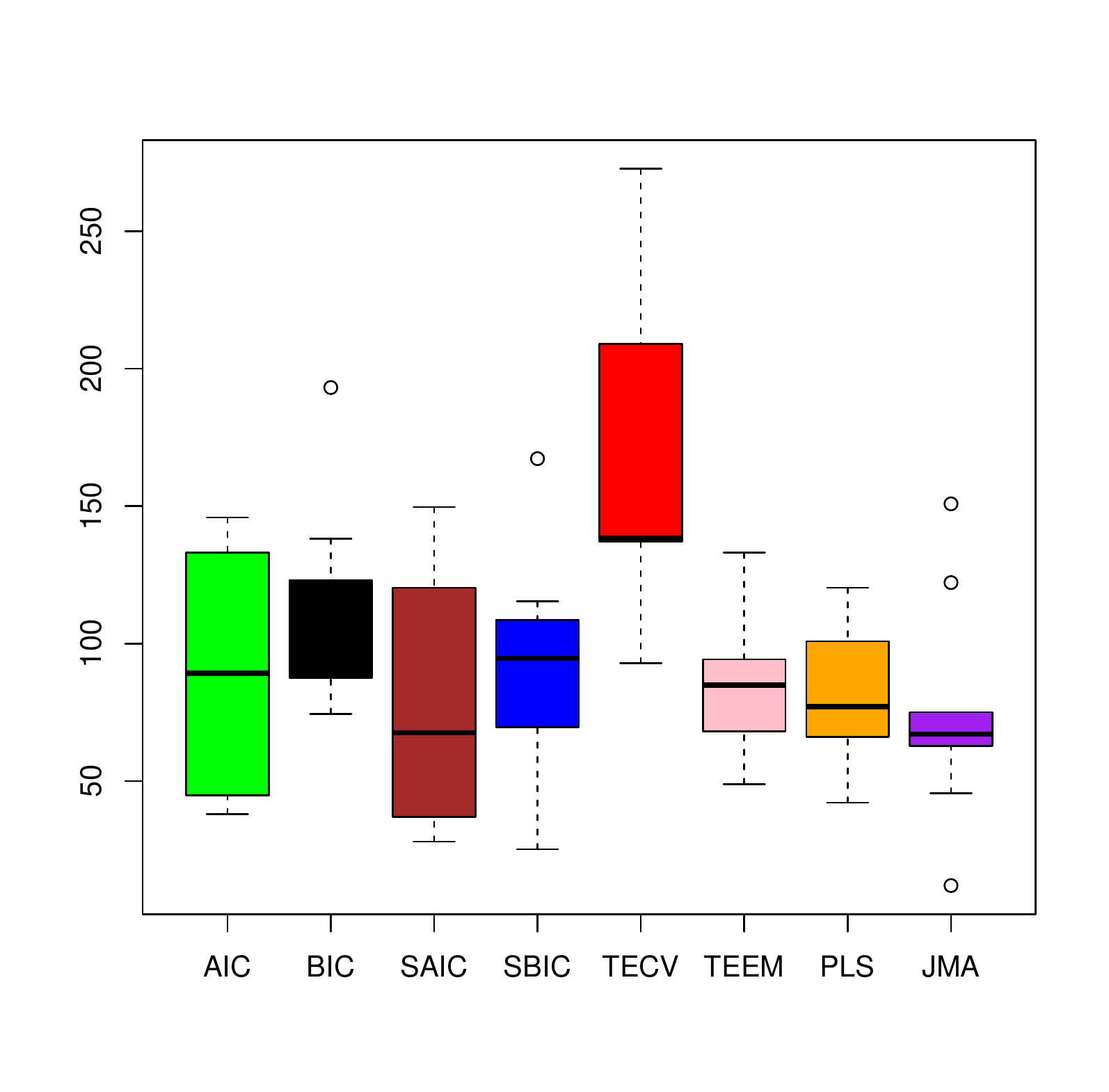}
 \vspace{-0.5cm}
	\caption{Box-plot of average squared errors in empirical application when we use the 19-th candidate model, which is chosen by AIC as the “true” process.}
 \vspace{-0.7cm}
\end{figure}

\section*{Acknowledgments}
\noindent{The authors are listed in alphabetical order, indicating equal contribution to the work. We would like to thank the anonymous reviewer for the helpful comments. This research was supported by National key R\&D Programmes of China (2022YFA1003800), National Statistical Science Research Grants of China (2022LD08), the National Natural Science Foundation of China (12231011, 71988101, 71925007 and 12288201), the CAS Project for Young Scientists in Basic Research (YSBR-008).}

\normalem
\bibliographystyle{Chicago}
\bibliography{ref}
\end{document}